\documentstyle[12pt,psfig]{article}
\newcommand{\be}{\begin{equation}}
\newcommand{\ee}{\end{equation}}
\begin{document}
\title{Coulomb Effects on Electromagnetic Pair Production 
in Ultrarelativistic Heavy-Ion Collisions}
\author{U.~Eichmann, J.~Reinhardt, W.~Greiner}
\date{Institut f\"ur Theoretische Physik,\\
Johann Wolfgang Goethe-Universit\"at, Frankfurt am Main, Germany}
\maketitle
\begin{abstract}
We discuss the implications of the eikonal amplitude on the 
pair production probability in ultrarelativistic heavy-ion transits. In this
context the
Weizs\"acker-Williams method is shown to be exact in the ultrarelativistic
limit, irrespective of the produced  particles' mass.  
A new equivalent single-photon distribution is derived which correctly 
accounts for the
Coulomb distortions. As an immediate application, 
consequences for unitarity violation in
photo-dissociation processes in peripheral heavy-ion encounters are
discussed. 
\end{abstract}

\section{Introduction}
The $S$ matrix describing electron scattering at ultrarelativistic pointlike 
charges was shown to be determined by the 
gauge phase leading to the Dirac equation represented in the temporal gauge
\cite{Eichmann}. 
We found that it naturally exhibits the same form as the well known eikonal
expression, as is expected by Lorentz invariance. 

The gauge phase leading to the temporal gauge reads
\[
\phi(x)=e^{-i\int_{-\infty}^{t}A_0(x)dt'}
\]
The consideration of asymptotic states corresponds to sending the upper
bound of the integral to infinity. 
The infinite time integral  over the scalar part of the
electromagnetic potential in the exponent 
has to be understood as the principal value of the integral. 
It can be decomposed into a finite term and an infinite quantity,
expressing the familiar divergence of phases in Coulomb scattering. The
infinite term can be removed by a gauge transformation which converts the
Coulomb boundary conditions of the original problem into a modified 
short-range potential allowing for asymptotic plane wave solutions. 

The ultrarelativistic limit of the gauge transformed potential $A_0'(x)$ 
reads \cite{Aichelburg,Baltz3,Segev1,Eichmann}
\be
\label{potlim}
\lim_{\gamma \to \infty} A_0'(x) = Z\alpha \delta (z-t) \ln x_\perp^2 
\ee
and hence we obtain for the $S$ operator in coordinate space 
($\hat{\gamma}_-=\hat{\gamma}_0-\hat{\gamma}_3$ is the Dirac matrix structure of
the interaction) 
\be
\label{sopcoo}
S=e^{-iZ\alpha \ln x_\perp^2 }\hat{\gamma}_-
\ee
The obtained $S$ operator is a
unitary operatorb due to its conformity to the gauge phase. 
It agrees with the first term of the Magnus expansion of   
the time-evolution-operator \cite{Magnus}, 
since the considered gauge-transformed 
interaction was assumed to be compressed to infinitely short times. 

This result was proven to be of completely
perturbative nature \cite{Eichmann}.
Note, however, 
that the perturbative derivation did not require the explicit 
Fourier transform of the transverse part of (\ref{potlim}),
which is an
ill defined object. 
For that reason the deduction of the small-coupling limit 
($Z\alpha\to 0$) of (\ref{sopcoo}) in momentum space 
must be treated with special care. 

In a rigorous distributional sense 
it can be defined as the {\it weak limit} $\lambda\to 0$
\cite{Ferrari}
\be 
\label{ftlog}
\int d^2x_\perp e^{-i\vec{k}_\perp\vec{x}_\perp}\ln x_\perp^2
=\lim_{\lambda \to 0}4\pi\left(\frac{-1}{k_\perp^2+ \lambda^2}-\pi\delta^2
(k_\perp)\ln\left(\frac{\lambda^2}{\mu^2}\right)\right)
\ee
with $\lambda =
\omega/\gamma$, $\mu=2/e^C$.
'Weak limit' means 
that the limit $\lambda\to 0$ 
has to be taken after having integrated the expression with a test function. 
The second term on the RHS arises from the gauge transformation applied to
the potential and thus accounts for the Coulomb distortions. 

Discarding the second term and taking the limit $\lambda\to 0$ 
directly would 
accidentally yield an expression for the 
Fourier transform of (\ref{potlim}), being identical to 
the Fourier transform of the ungauged
potential in the limit $\gamma\to \infty$ 
\[                                      
\int d^4x \;e^{ikx}A_0=-(2\pi)^2 Z \alpha \delta (k_-)\frac{2}{k^2_\perp}
\]

This error is made if one intends to extract the correct small-coupling
limit from a naive Taylor expansion 
of the Fourier transformed $T$ matrix 
whose linear term reads
\be
\label{ttaylor}
\lim_{Z\alpha\to 0}T(k)\approx (2\pi)^2\delta (k_-) iZ \alpha
\frac{2}{k^2_\perp} \overline{u}(p')\hat{\gamma}_-u(p)
\ee
Here $u(p)$ is the electron unit spinor,
$p$ and $p'$ are the initial and final  momenta of the electron,
repectively, $k=p'-p$.

Since, however,  Taylor expansion and Fourier transformation 
do not commute in this case, 
the Taylor expansion of the Fourier transformed $T$ matrix for this purpose
is not justified.\footnote{Note, that with the above mentioned
distribution-theoretical precautions, it is possible to obtain the correct
result via Taylor expansion \cite{Grignani}.}
The correct small-coupling limit of the scattering amplitude in
momentum space can thus not be found by a naive 
Taylor expansion of the Fourier transformed $T$ matrix and does not agree 
with first-order
perturbation theory. According to (\ref{ftlog}) this is simply 
based on the fact, that 
the gauge transformed potential correctly
accounts for Coulomb boundary conditions. 

In the following 
we want to investigate, how the correct treatment of Coulomb boundary
conditions in all orders of 
perturbation theory influences the cross section of
the scattering process.

\section{Implications on the cross section}

We consider the exact amplitude for electron scattering at
an ultrarelativistic pointlike charge, moving in $+z$ direction. 
We state it in terms of the invariant squared momentum transfer 
$t\approx-k_\perp^2$ for
$\gamma\to \infty$\footnote{Note the striking
similarity between (\ref{ultraamp}) and the nonrelativistic (Schr\"odinger) 
amplitude
\[
f(\theta)=-\frac{1}{2k^2\sin^2\frac{\theta}{2}}\frac{\Gamma
\left(1+\frac{i}{k}\right)}{\Gamma
\left(1-\frac{i}{k}\right)}e^{-\frac{i}{k}\ln \sin^2 \frac{\theta}{2} }
\]
with the squared momentum transfer being proportional to $\sin^2 \theta /2$.
}
\begin{eqnarray}
\label{ultraamp}
A&=& 2\pi\delta (k_-) F_{p',p}(e^{-iZ\alpha \ln x_\perp^2}-1)
\overline{u}(p')\hat{\gamma}_-u(p)\nonumber \\
&=&-i8 \pi^2 Z\alpha
\delta (k_-)\frac{1}{t}
\frac{\Gamma(-i\alpha Z)}{\Gamma(i\alpha Z)}
e^{-iZ\alpha \ln (-t/4)}\overline{u}(p')\hat{\gamma}_-u(p)
\end{eqnarray}
$F_{p',p}$ abbreviates the Fourier transform with respect to the transverse
coordinates, taken at the momentum 
$\vec{k}_\perp=(\vec{p'}_\perp-\vec{p}_\perp)$.
The cross section for this scattering process is found to be exactly the Mott
formula for Coulomb scattering of ultrarelativistic electrons 
at a static source, 
Lorentz-transformed into the electron's rest frame. 
Such kind of agreement between the exact result and the first order
perturbation theory is also found in the nonrelativistic case, known
as one of the peculiarities of the Coulomb field.

The well established 
eikonalization of the scattering amplitude and thus the 
reduction to Mott's result imply, that in the high-energy limit 
the electron and the positron Coulomb scattering cross section
become identical. This behaviour of the cross section at
ultrarelativistic energies 
is required by the Pomeranchuk theorem \cite{Itzykson}. 

One can draw an analogy to pomeron exchange in hadron physics: 
The scattering process can be described in terms of the single exchange of
an ''effective photon`` according to the modified potential
\be
\label{modpot}
V_0(x)= V_3(x)
=\delta (z-t)\left(\left(\frac{1}{x_\perp}\right)^{2iZ\alpha}-1\right)
\ee

In the field of
two ultrarelativistic colliding pointlike nuclei, 
the exact scattering amplitude 
was shown to retain the structure of the
second-order perturbative result, due to the causal decoupling property
\cite{Eichmann}. 
Each interaction can be described by the modified potential (\ref{modpot}).
In the following we consider the symmetric collision of two 
ions with charge $Ze$, the extention to asymmetric collisions is trivial. 

Accounting for both time orderings, the amplitude reads
\begin{eqnarray}
A^{tot}_{p'p}&=&\int \frac{d^2k_\perp}{(2\pi)^2}
F_{k,p}(e^{-iZ\alpha \ln x_\perp^2}-1)
F_{p',k}(e^{-iZ\alpha \ln x_\perp^2}-1)
e^{i(\vec{p'}_\perp-\vec{k}_\perp)\cdot \vec{b}}
\nonumber \\
&&\left(\overline{u}(p')\frac{-\hat{\vec{\alpha}}_\perp 
\cdot \vec{k}_\perp + 
\gamma_0
m}{p'_+ p_- - {k}_\perp^2 -m^2 +i\epsilon} 
\hat{\gamma}_+u(p)\right.
\nonumber\\
\label{Ttot}
&&\left.\overline{u}(p')\frac{-\hat{\vec{\alpha}}_\perp
\cdot (\vec{p}_\perp +\vec{p'}_\perp -\vec{k}_\perp) +
\gamma_0
m}{p'_- p_+ -
(\vec{p}_\perp +\vec{p'}_\perp -{k}_\perp)^2 -m^2 +i\epsilon}
\hat{\gamma}_-u(p)\right)
\end{eqnarray}
Here the trajectory of one ion is shifted by the impact
parameter $\vec{b}$, which is accounted for by the factor
$e^{i(\vec{p'}_\perp-\vec{k}_\perp)\cdot \vec{b}}$.

We now use the crossing invariance of the amplitude to apply the
obtained result to electron-positron pair production. The initial electron
four momentum $p$ has then to be replaced by the negative final positron
momentum $p\to -p^p$. The final electron momentum will be
denoted by $p'=p^e$.
With (\ref{Ttot}) we obtain for the pair production probability
\begin{eqnarray}
\frac{d\sigma}{d^2b} &=&  |A^{tot}_{p'p}|^2
\frac{md^3p^e}{(2\pi)^3E^e}\frac{md^3p^p}{(2\pi)^3E^p}
\nonumber \\
&=&
\frac{md^3p^e}{(2\pi)^3E^e}\frac{md^3p^p}{(2\pi)^3E^p}
\int \frac{d^2k_\perp}{(2\pi)^2}\int \frac{d^2k'_\perp}{(2\pi)^2}
F_{k,-p^p}(e^{-iZ\alpha \ln x_\perp^2}-1)
F_{p^e,k}(e^{-iZ\alpha \ln x_\perp^2}-1)\nonumber\\
&&F^\ast_{k',-p^p}(e^{-iZ\alpha \ln x_\perp^2}-1)
F^\ast_{p^e,k'}(e^{-iZ\alpha \ln x_\perp^2}-1)
e^{i(\vec{k'}_\perp-\vec{k}_\perp)\cdot \vec{b}}\nonumber
\\
&&\left(\overline{u}(p^e)\frac{-\hat{\vec{\alpha}}_\perp
\cdot \vec{k}_\perp +
\gamma_0
m}{-p^e_+ p^p_- - {k}_\perp^2 -m^2 +i\epsilon}
\hat{\gamma}_+u(-p^p)\right.\nonumber\\
&&\left.
+\overline{u}(p^e)\frac{-\hat{\vec{\alpha}}_\perp
\cdot (-\vec{p^p}_\perp +\vec{p^e}_\perp -\vec{k}_\perp) +
\gamma_0
m}{-p^e_- p^p_+ -
(-\vec{p^p}_\perp +\vec{p^e}_\perp -{k}_\perp)^2 -m^2 +i\epsilon}
\hat{\gamma}_-u(-p^p)\right)\nonumber\\
&&\times\left(\overline{u}(p^e)\frac{-\hat{\vec{\alpha}}_\perp
\cdot \vec{k'}_\perp +
\gamma_0
m}{-p^e_+ p^p_- - {k'}_\perp^2 -m^2 +i\epsilon}
\hat{\gamma}_+u(-p^p)\right.\nonumber\\
\label{csb}
&&\left.
+\overline{u}(p^e)\frac{-\hat{\vec{\alpha}}_\perp
\cdot (-\vec{p^p}_\perp +\vec{p^e}_\perp -\vec{k'}_\perp) +
\gamma_0
m}{-p^e_- p^p_+ -
(-\vec{p^p}_\perp +\vec{p^e}_\perp -{k'}_\perp)^2 -m^2 +i\epsilon}
\hat{\gamma}_-u(-p^p)\right)^\ast
\end{eqnarray}

The integration over the impact parameter yields the pair production cross
section. 
Due to the $\delta^2(\vec{k'}_\perp-\vec{k}_\perp)$-function occuring in 
the $\vec{b}$ integration, 
a further momentum integral can be perfomed, leaving 
\begin{eqnarray}
d\sigma&=&\frac{md^3p^e}{(2\pi)^3E^e}\frac{md^3p^p}{(2\pi)^3E^p}\int
\frac{d^2k_\perp}{(2\pi)^2}
|F_{k,-p^p}(e^{-iZ\alpha \ln x_\perp^2}-1)|^2
|F_{p^e,k}(e^{-iZ\alpha \ln x_\perp^2}-1)|^2\nonumber
\\
&&\left|\overline{u}(p^e)\frac{-\hat{\vec{\alpha}}_\perp
\cdot \vec{k}_\perp +
\gamma_0
m}{-p^e_+ p^p_- - {k}_\perp^2 -m^2 +i\epsilon}
\hat{\gamma}_+u(-p^p)\right.\nonumber\\
\label{cs}
&&\left.+\overline{u}(p^e)\frac{-\hat{\vec{\alpha}}_\perp
\cdot (-\vec{p^p}_\perp +\vec{p^e}_\perp -\vec{k}_\perp) +
\gamma_0
m}{-p^e_- p^p_+ -
(-\vec{p^p}_\perp +\vec{p^e}_\perp -{k}_\perp)^2 -m^2 +i\epsilon}
\hat{\gamma}_-u(-p^p)\right|^2
\end{eqnarray}
Thus, upon integration over the whole impact parameter plane, 
the phases in the individual scattering amplitudes (see (\ref{ultraamp})) 
cancel. Consequently, in the limit $\gamma\to \infty$ the 
cross section is found to reduce 
to the lowest-order
perturbation theory, 
the two-photon result. 
This behaviour does not only naturally explain \cite{Segev2} 
the experimentally 
observed quadratic
dependence on the target's and the projectile's charge 
\cite{Vane}, 
but also implies,
that no asymmetries should occur in 
electron and positron spectra.  

Equation (\ref{cs}) is strictly valid only for pointlike ions. 
The focus on
electromagnetic reactions in peripheral heavy-ion collisions 
implies a restricted range of impact parameters with a lower bound 
$b=r_A+r_B$, $r_A$ and $r_B$ being the radii of the ions.
Therefore 
in experiments which are triggered on peripheral collisions, effects of the 
Coulomb distortions 
described 
by the phase in (\ref{ultraamp}) will be visible. 

The eikonal approximation (and thus the cross section) is known to become
energy-independent in the ultrarelativistic limit \cite{Cheng-Wu}. 
This dependence  
is restored by accounting for the correct transverse 
momentum range, which is restricted by the validity of (\ref{potlim}). This
condition reads \cite{Baltz,Eichmann}
\be
\label{applcond}
|\vec{k}_\perp|\gg \frac{\omega}{\gamma}
\ee
Such a low energy cut off is also necessary to cure the divergence in
(\ref{Ttot}). 

\section{Equivalent Photon Approximation}
We intend to study the behaviour of the cross section,
both impact parameter dependent and impact parameter integrated, in the
Weizs\"acker-Williams method of equivalent photons. 
This approximation uses the similarity between the fields of a fast
moving charge and a swarm of real photons moving in beam direction. It
approximately correstponds to the first order Born approximation in the
temporal gauge: Only the transverse part of the interaction is considered -- 
the longitudinal part is suppressed by $1/\gamma^2$ -- and the vertex
function is evaluated on-shell at $k^2=0$, i.e. for an assumed real photon.
Rewriting the exact  
cross section in terms of the real photon cross section, the
whole information about the scattering potential, which can be the retarded
Coulomb potential or the modified potential (\ref{modpot}), respectively, is
then contained in the distribution function of the equivalent photons
$n(\omega)$. 
Roughly speaking, this photon distribution function is determined by the
squared absolut value of the Fourier-transformed potential (in temporal
gauge).
The obvious 
advantage of casting the exchange of effective photons according to
(\ref{modpot}) in the Weizs\"acker-Williams form is, that 
any difference between the second-order
perturbative result and the exact calculation will be solely generated by 
differences between the equivalent photon distributions.

The Weizs\"acker-Williams approximation is applicable if the exchanged
momentum meets the following conditions \cite{LandauIV} 
\be
\label{wwbed1}
\frac{\omega}{\gamma^2}\ll |\vec{k}_\perp|\ll m
\ee
and
\be
\label{wwbed2}
|\vec{k}_\perp|\ll \omega \ll m \gamma
\ee
The upper bounds mainly stem from the requirement, that $|k^2|$
 is negligible compared to $2p\cdot k \ge m^2$, such that the intermediate
 propagators of the scattered particles 
can be approximated by those describing the interaction with real photons. 
The particle's rest mass in (\ref{wwbed1}) is however 
a conservative upper
bound and the equivalent photon method is not strictly invalid for $|k^2|
\sim m^2$. 
Note, that for the approximative
calculations in \cite{Bottcher}, the transverse
mass of the scattered particle was taken as the upper bound for
(\ref{wwbed1}).

Since the exact amplitude takes the eikonal form, we must point out the
following: 
The expansion of the ultrarelativistic
scattering amplitude in powers of the transferred
momentum yields, as the leading term, the eikonal expression (describing the
minimal deflection from the initial straight-line trajectory)
\cite{Abarbanel,Sugar}. Its
perturbation-theoretical derivation requires that the
quadratic terms $k^2$ are negligible 
relative to the terms $2p_i\cdot k$ in the denominators
of the propagators, where $k$ is any partial sum of the internal momenta
\cite{Levy}.
The exact validity of the eikonal formula at infinite energies 
therefore shows, that the 
transferred momentum $|k^2|$ does not exceed $m^2$ (irrespective of the
value
of $m$). 
This agrees with the theoretical observation, that at high energies
particles
are predominantly scattered into a cone with opening angle $\theta \sim 1/
\gamma$, corresponding to momentum transfers $|k^2|\sim m^2$. The main
contributions to the cross section are thus expected from spatial distances
larger or equal the Compton wavelength of the particle.\footnote{From the
asymmetry of electron and positron spectra
produced in $S$(200 GeV/n)+Au collisions , the
mean transverse distance from the target ion was deduced to be approximately
two Compton wavelengths \cite{Vane}. The collision energy corresponds to
$\gamma \approx 10$ in the center of speed system.}

Moreover, the longitudinal part of the interaction vanishes identically in
the
limit $\gamma\to \infty$.
Hence, the applicability conditions of the Weizs\"acker-Williams
approximation
are trivially fulfilled in the limit $\gamma\to \infty$.\footnote{Just as
the
eikonal formula, the Weizs\"acker-Williams approximation can be viewed as
the leading term of an expansion in powers of $k^2/m^2$ \cite{Brodsky}. The
validity of the eikonal expression then automatically implies the validity
of
the Weizs\"acker-Williams method.}

\begin{figure}[h]
\begin{center}
\begin{minipage}{16cm}
\parbox{7cm}{\psfig{figure=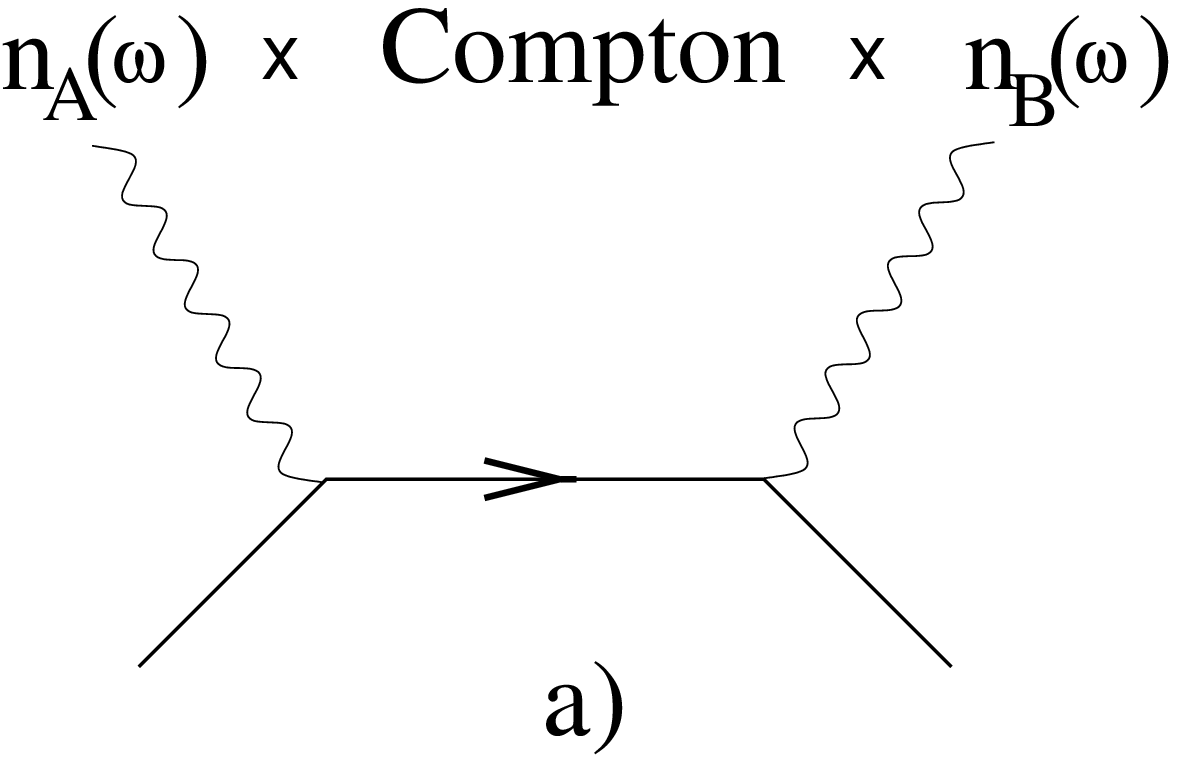,width=6cm}}\hspace{1cm}
\parbox{7cm}{\psfig{figure=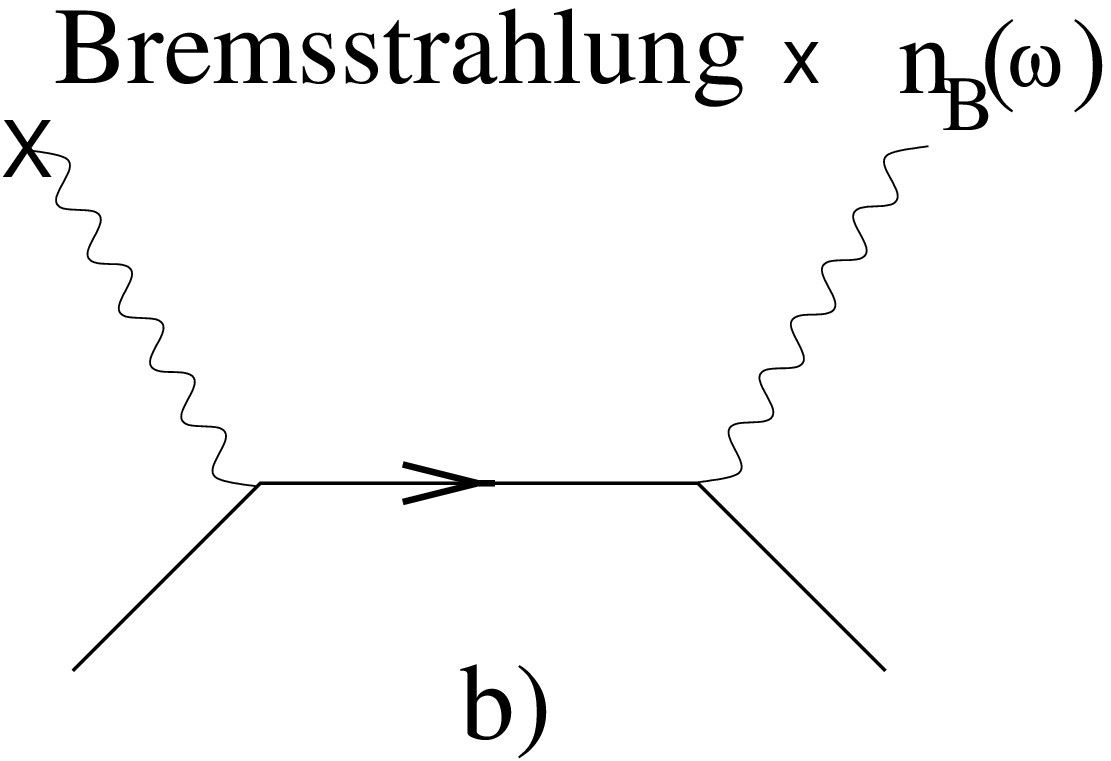,width=6cm}}
\end{minipage}
\end{center}
\caption{\label{fig01}The two possible distinct processes, that can be used
to describe electron-positron production in heavy-ion collision. One or both  
ultrarelativistic ions can be replaced by an equivalent photon distribution.
If the bremsstrahlung process b) is calculated in one ion's rest frame, the
electron must be assumed ultrarelativistic, to yield agreement with
a).}
\end{figure}

We have the freedom to apply this method to the interaction of the electron
with both nuclei, 
giving two possibilities (see Figure \ref{fig01})

The two possible calculation schemes (Figure \ref{fig01})
agree, since in b) 
the bremsstrahlung emission and the scattering at the external
potential decouple. This is due to the fact, that the region in which the
ultrarelativistic electron ''feels`` the external field is assumed to be pointlike and
any frequency of the emitted photon is ''soft`` compared to the timescale of
the scattering.
Coulomb effects arise, if one explicitly accounts for the finite
interaction time, either in the scattering process by correcting the eikonal
formula
\label{eikexp} 
or by keeping the eikonal amplitude for the scattering process
but assuming a Rutherford-deflected trajectory for the photon emission
\cite{LandauIV}. Corrections to the eikonal formula account for higher
orders of e.g. the Magnus expansion \cite{Magnus},
which is an expansion in the interaction time $\tau$
around the
instantaneous interaction $\tau\sim 1/\gamma\to 0$.
In general these Coulomb effects vanish, if the
energy of the emitted photon is too small to resolve details of the
scattering process, and the recoil of the electron is negligible.  

To apply the Weizs\"acker-Williams method to the bremsstrahlung photon, 
the recoil of the
bremsstrahlung photon
must, however, be assumed
negligible. 
The small momentum
transfer 
is in turn ensured by the
eikonalization of the scattering process. 

The equivalent single-photon distributions $n_{A/B}(\omega)$ 
of the ions $A$
and $B$, are determined from the effective potential
(\ref{modpot}). 
The photon distribution reads 
\be
n(\omega)=\frac{1}{4\pi^2 \alpha \omega} 
\int_{\omega/\gamma}^{m} k_\perp dk_\perp 
\left|k_\perp\pi
Z\alpha\left(\frac{4}{k_\perp^2}\right)^{1-iZ\alpha}\frac{\Gamma(-i\alpha
Z)}{\Gamma(i\alpha Z)}\right|^2=
\frac{2Z^2 \alpha}{\pi\omega}
\ln\left(\frac{m\gamma}{\omega}\right)
\ee
The lower bound of the integral is taken from the condition
(\ref{applcond}).
The upper bound, the electron rest mass, is imposed by (\ref{wwbed1}).
The prefactor arises from properly rewriting the cross
section (\ref{cs}) in terms of the real photon cross section (i.e. the
Compton cross section) and photon
distribution functions. 

This photon distribution 
coincides with the equivalent-photon distribution obtained from the
Coulomb potential to logarithmic accuracy \cite{LandauIV} 
and is not changed by Coulomb
effects.

\section{Impact parameter dependent cross section}
The impact parameter dependent equivalent photon method for the exact
calculation, using the modified potential (\ref{modpot}), 
can be derived similarly to
\cite{Vidovic}. 
To this end 
we have to modify the integrands in (\ref{csb}) such that they  
account for the limited momentum range ((\ref{applcond}) and
(\ref{wwbed1})).

The cut off of low transverse momenta according to (\ref{applcond}) can be
achieved by the following replacement in 
(\ref{ultraamp})\footnote{Note, that for the
Schr\"odinger case the exact validity of the eikonal formula 
can be proven for a certain off shell domain of the 
momentum transfer for the whole energy plane \cite{Banyai}.} 
\be
\label{text}
t=-k_\perp^2 \to -\frac{\omega^2}{\gamma^2}-k_\perp^2
\ee
This substitution suppresses small transverse momenta less strongly than the
strict cut off at $k_\perp=\omega /\gamma$. It assumes the sufficient
accuracy of the classical
eikonal amplitude for
$1/\gamma$ in the near vicinity of $1/\gamma =0$, which is guaranteed by the
possibility of continuous extentions of the eikonal formula towards finite
$\gamma$ and large $t$. 
A Yukawa-type damping of transverse distances corresponding to the cut off
transverse momenta yields additional terms that change the character of the
amplitude significantly and can not be motivated physically. 

In accordance with the exact validity of the eikonal formula, the physical
situation is such, that the transferred momenta are restricted by  the
condition $|k^2|\ll m^2$. They are, however, 
naturally cut off, if one introduces 
a form factor to account
for the finite extent of the nuclei. 
Thus, large momenta 
have to be cut off at $k_\perp\approx 1/r_>$, where $r_>$ is the larger 
value of either 
the nuclear radius or the Compton wavelength of the scattered
particle \cite{Jackson}. In this respect, the electron is an
exception, since all other particles have Compton wavelengths smaller or
comparable to the nuclear size.
To present the calculations in a uniform manner, we use the form factor of
the nucleus to cut off the large momenta. 

The impact parameter dependent cross section for particle production,
described in the equivalent photon method reads \cite{Vidovic}
\be
\label{csbww}
\frac{d\sigma}{d^2b}=\int d\omega_1 \int d\omega_2
\left[n_\|(\omega_1,\omega_2,\vec{b})
\sigma^{\gamma\gamma}_\|(\omega_1,\omega_2) +
n_\perp(\omega_1,\omega_2,\vec{b})
\sigma^{\gamma\gamma}_\perp(\omega_1,\omega_2)\right]
\ee
with the two-photon distribution functions
$n_{\|/\perp}(\omega_1,\omega_2,\vec{b})$. The elementary two-photon cross
sections and the two-photon distribution functions explicitly account for
the parallel or orthogonal orientation of the photon-polarizations, denoted
by the indices $\|$ and $\perp$, respectively. 
Since the integration over the impact
parameter plane implies an averaging over the photon polarizations, the
explicit occurrence of the photon polarizations in the impact parameter
dependent cross section is expected.
The functions
$n_{\|/\perp}(\omega_1,\omega_2,\vec{b})$ can be expressed in terms of
single-photon distribution functions $n(\omega,b)$, 
depending on the transverse separation:
\begin{eqnarray}
\label{n2par}
n_\|(\omega_1,\omega_2,\vec{b})&=&\int d^2x_\perp
n(\omega_1,\vec{x}_\perp-\vec{b})\,n(\omega_2,\vec{x}_\perp)
\left(\frac{(\vec{x}_\perp-\vec{b})\cdot
\vec{x}_\perp}{|\vec{x}_\perp-\vec{b}||\vec{x}_\perp|}\right)\\
\label{n2senk}
n_\perp(\omega_1,\omega_2,\vec{b})&=&\int d^2x_\perp
n(\omega_1,\vec{x}_\perp-\vec{b})\,n(\omega_2,\vec{x}_\perp)
\left(\frac{(\vec{x}_\perp-\vec{b})\times 
\vec{x}_\perp}{|\vec{x}_\perp-\vec{b}||\vec{x}_\perp|}\right)
\end{eqnarray}
with
\be 
\label{photdist}
n(\omega,b)=\frac{Z^2\alpha}{\pi^2\omega}\left|\int_0^\infty dk_\perp k_\perp^2
\frac{F(k_\perp^2+\omega^2/\gamma^2)}{(k_\perp^2+\omega^2
/\gamma^2)^{1-iZ\alpha}}J_1(bk_\perp)\right|^2
\ee
$J_1$ is a Bessel function. The function $F$ denotes the chosen form
factor of the nucleus. 

For a pointlike charge ($F\equiv 1$), the photon distribution function can
be calculated analytically. We obtain for the Coulomb potential and the
modified potential 
\be
\label{photdistpt}
n(\omega,b)=\left\{\begin{array}{ll}
\displaystyle \frac{Z^2\alpha}{\pi^2}\frac{\omega}{\gamma^2}\left[K_1\left(\frac{\omega
b}{\gamma}\right)\right]^2\;\;\;&{\rm retarded\;Coulomb\;potential}\\
&\\
\displaystyle
\frac{Z^2\alpha}{\pi^2}\frac{\omega}{\gamma^2}\left[\frac{K_{1+iZ\alpha}(\omega
b/\gamma)}{\Gamma(1-iZ\alpha)}\right]^2&{\rm
modified\;potential\;(\ref{modpot})}
\end{array}\right.
\ee
$K_\nu$ is a modified Bessel function. For small arguments of the Bessel
function one can use the asymptotic expression \cite{Abramowitz}
\be
\label{kasym}
K_\nu (z)\sim \frac{1}{2}
\Gamma (\nu )
(\frac{1}{2} z)^{-\nu }
\ee
Therefore, for $\omega b \ll \gamma$ and assumed
point-like charges the photon
distribution functions (\ref{photdistpt}) nearly completely agree. 
\newpage 
\begin{figure}[h]
\centerline{\psfig{figure=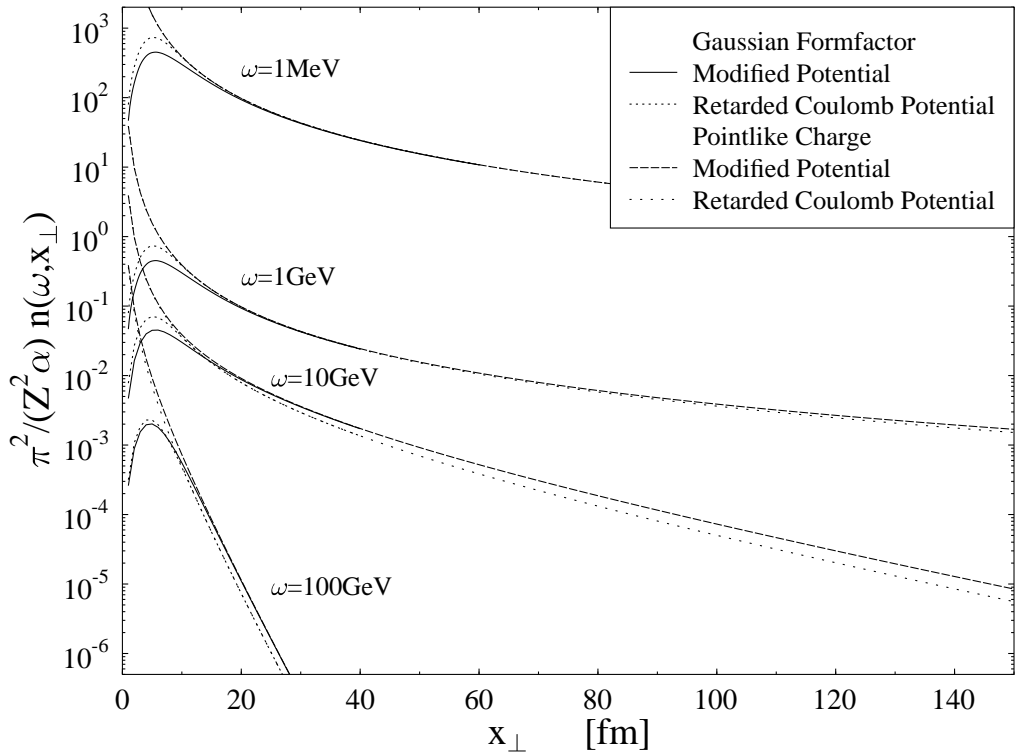,width=17cm}}
\caption{\label{fig02}The single-photon distribution function for various 
photon energies (indicated in the plot) as
a function of the transverse distance from the ultrarelativistic charge. The
calculations are done for lead ions ($Z=82$) at
LHC energies ($\gamma\approx 3000$). }
\end{figure}

\begin{figure}[hbtp]
\centerline{\psfig{figure=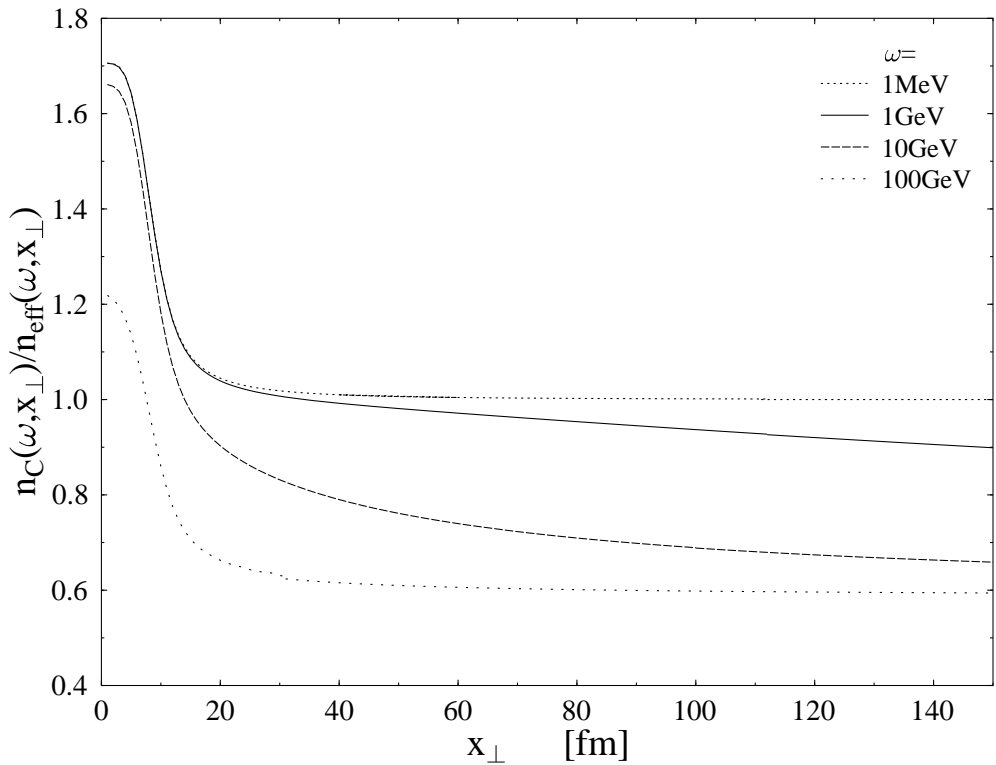,width=17cm}}
\caption{\label{fig03}The ratio of the equivalent photon numbers of the
pure Coulomb potential, $n_C(\omega,x_\perp)$, and the photon numbers of the
effective potential, $n_{eff}(\omega,x_\perp)$. At small transverse
distances, one finds a deviation of up to $70\%$ for small photon energies.
Far outside the nucleus, the photon distribution functions are determined by
(\protect\ref{photdistpt}). For large distances one asymptotically finds a
deviation in the order of $40\%$, independent of the photon energy.}
\end{figure}

We have numerically evaluated the photon distribution function
(\ref{photdist}) for an extended nucleus, using a gaussian form factor
$F(Q^2)=e^{-Q^2/(2Q_0^2)}$ with $Q_0=60MeV$ which describes the $Pb$ nucleus 
\cite{Drees}. 
Figure \ref{fig02} shows a comparison between the photon distribution
functions for both, point like nuclei ($F\equiv 1$) and extended nuclei
usin either 
the
retarded Coulomb potential or the modified potential (\ref{modpot}),
which represents the exact calculation. 
At small distances up to a few multiples
of the nuclear radius, the modified potential gives a smaller number of
equivalent 
photons than the pure Coulomb potential. For large distances or large photon
energies (\ref{kasym}) looses its validity and the modified potential gives
a larger number of photons than the Coulomb potential (see Figure
\ref{fig03}). 

The significance of these results can be assessed 
in single-photon induced processes,
such as the electromagnetic dissociation of nuclei in peripheral heavy-ion
collisions. In the Weizs\"acker-Williams approach 
the dissociation probability at a given impact parameter is
calculated by integrating 
the product of the measured 
photon-nucleus dissociation cross section of the target,
$\sigma_T^\gamma(\omega)$, 
and the
equivalent photon distribution, $n_P(\omega,b)$, 
of the projectile (\ref{photdist}) 
over the photon energy \cite{Vidovic2,Norbury0}
\[ 
P(b)=\int d\omega \,n_P(\omega,b)\, \sigma_T^\gamma(\omega)
\]
As shown in \cite{Vidovic2} for $Pb+Pb$ collisions at LHC energies, 
the dissociation 
probability violates unitarity at small impact parameters up to 25 fm. At
small impact parameters, however, the photon distribution of the modified
potential is suppressed relative to that of the pure Coulomb potential and
yields a reduction of the probabiltiy, whereas at large impact parameters,
the dissociation probability is enhanced. The reduction visible in Figure
\ref{fig02} taken alone is too small to cure the unitarity violation.
Possibly this can be achieved if in addition 
multi-phonon excitations \cite{Bertulani,Norbury1} are taken into account,  
which also reduce the dissociation  probability at small impact parameters. 

The impact parameter dependent two-photon distribution functions also have
to be corrected according to (\ref{photdist}), thus correcting the
pair production probability (\ref{csbww}), i.e. (\ref{csb}), for Coulomb
effects. Due to the complicated convolution of the single-photon
distribution functions in (\ref{n2par}) and (\ref{n2senk}), effects on the
two-photon distribution function are not obvious. For large distances
$x_\perp \gg b$, however, one directly finds an enhancement of the
equivalent two-photon numbers.

\section{Summary}
It was shown by either 
summing the perturbation series \cite{Eichmann} or by
matching plane waves at the delta function potential on the light front
\cite{Segev1,Baltz,Eichmann}, that the eikonal expression for the scattering
amplitude becomes exact in the ultrarelativistic limit ($\gamma\to \infty$).
This allows to neglect the squared momentum transfer $k^2$ 
relative to the term $2p_i\cdot k$ in the denominator of the propagator of the
scattered particle. As a consequence 
the applicability
conditions of the Weizs\"acker-Williams method are fulfilled automatically 
-- irrespective of the mass of the scattered particle. 

Furthermore, the exact validity of the eikonal formula for ultrarelativistic
scattering processes confirm the Pomeranchuk theorem,
stating that
the cross sections for antiparticle and particle scattering at a given
target become identical in the ultrarelativistic limit. In analogy to
pomeron exchange in hadron physics, on can describe the exact interaction as
the exchange of an effective photon, according to a modified, effective
potential given by (\ref{modpot}). 
The cross section, as a peculiarity of the Coulomb interaction, becomes
identical to the Mott result. The exact 
pair production cross section in the field
of two ultrarelativistic colliding (pointlike) ions also reduces to the
second order perturbative result \cite{Baltz} which was evaluated in 
\cite{Bottcher}. This allows for two
conclusions: i) The production rate scales with the square of the target and
the projectile charge \cite{Vane,Segev2} ii) Asymmetries in the electron and
positron spectra should not occur. 

Note, however, that the presented formalism is valid only if the produced
particles are fas with respect to both nuclei. 
Therefore, the observed \cite{Vane} asymmetry at small electron and 
positron momenta remains unaffected by these considerations. 

We applied the Weizs\"acker-Williams approach to pair production using the
modified potential (\ref{modpot}), correctly accounting for the Coulomb
boundary conditions. The impact parameter dependent single-photon
distribution calculated with the modified potential shows 
deviations from the equivalent photon distribution function obtained from
the retarded Coulomb potential in order of up to $70\%$ at small separations 
and approximately $40\%$ at large separations from the ion. 

In combination with multi-phonon excitation, the correct treatment of
Coulomb distortions possibly solves the problem of unitarity violation in
photodissociation processes in ultrarelativistic heavy-ion collisions. 

The pair production probability is also subject to changes due to the
modified photon numbers at given impact parameters and photon energies. 
The perturbation theoretical probability, as calculated here, rather
represents the average number of produced pairs and exceeds unity at
sufficiently small
impact parameters. The ''true`` pair
production probability has to be corrected by the vacuum-to-vacuum
amplitude, which in turn can be calculated from the perturbative pair
production probability \cite{Best,Rhoades-Brown}. This nontrivial influence
on the pair production cross section is subject of further studies. 

\section*{Acknowledgements}
This work was supported by
{\it Deutsche Forschungsgemeinschaft} DFG within the project Gr-243/44-2.

\end{document}